\documentclass[11pt]{article}

\usepackage{graphicx} % Required for inserting images
\usepackage[utf8]{inputenc}
\usepackage{eurosym,geometry,graphicx,color,setspace,sectsty,footmisc,natbib,pdflscape,array,hyperref}
\usepackage[normalem]{ulem}
\usepackage{caption}
\usepackage{subfigure}
\usepackage{amssymb,amsfonts,amsmath}
\usepackage{amsthm}
\usepackage{natbib}
\usepackage{booktabs}
\usepackage[flushleft]{threeparttable}
\usepackage{tikzsymbols} % emoji
\usepackage{tikz}
\usetikzlibrary{patterns,arrows}
\usepackage{apxproof}
\usepackage{siunitx}

\usepackage{xcolor} % Colored text
\usepackage{todonotes}
\setuptodonotes{inline}
\usepackage[]{mdframed}

\theoremstyle{definition} % Upright text
\newtheorem{definition}{Definition}

\newtheorem{example}{Example}
\newtheorem*{example*}{Example}

\newtheorem*{fact*}{Fact}

\AtBeginEnvironment{example}{%
  \pushQED{\qed}%
}
\AtEndEnvironment{example}{\popQED\endexample}

\usepackage[final]{changes} % Use this to NOT show changes in red
\definechangesauthor[color=red, name={Paul S. Koh}]{PSK}

\hypersetup{colorlinks,citecolor=blue} % citation in blue
\title{Market Definition: A Sensitivity Analysis\footnote{I thank Louis Kaplow, Sam Kleiner, Matt Panhans, Devesh Raval, Ted Rosenbaum, and David Schmidt for providing comments. \added[id=PSK]{I also thank the editor and two anonymous referees for their help in improving this paper.} \replaced[id=PSK]{Much of this article was written during the author’s tenure at the U.S. Federal Trade Commission; the views expressed in this article are those of the author and do not necessarily represent those of the Federal Trade Commission or any of its Commissioners.}{The views expressed in this paper are those of the author. They do not necessarily represent those of the Federal Trade Commission or any of its Commissioners.} \added[id=PSK]{The author declares no conflict of interest. During the preparation of this work, the author used ChatGPT (OpenAI’s language model) in order to check grammar and improve the clarity of the writing. After using this tool/service, the author reviewed and edited the content as needed and takes full responsibility for the content of the publication.} All errors are mine. \deleted[id=PSK]{I welcome any comments.}}}
\author{Paul S. Koh\footnote{\replaced[id=PSK]{School of Economics, Yonsei University. 50 Yonsei-ro, Seodaemun-gu, Seoul, Republic of Korea. Email: \texttt{paulkoh9@gmail.com}.}{Federal Trade Commission, 600 Pennsylvania Avenue NW, Washington, DC 20580. Email: \texttt{pkoh@ftc.gov.}}}}
\date{August 24, 2025}

\begin{document}

\maketitle
\begin{abstract}
Market definition holds significant importance in antitrust cases, yet achieving consensus on the correct approach remains elusive. As a result, analysts routinely entertain multiple market definitions to ensure the resilience of their conclusions. I propose a simple framework for conducting organized sensitivity analysis with respect to market definition. I model candidate market definitions as partially ordered and use a Hasse diagram, a directed acyclic graph representing a finite partial order, to summarize the sensitivity analysis. I use the Shapley value and the Shapley-Shubik power index to quantify the average marginal contribution of each firm in driving the conclusion. I illustrate the method's usefulness with an application to the Albertsons/Safeway (2015) merger.

    \vspace{1em}
\end{abstract}

\clearpage

\section{Introduction}

Over the course of U.S. antitrust litigation history, the resolution of a greater number of cases has hinged on \emph{market definition}, more so than any other significant aspect \citep{baker2007market}. Courts in the United States and Europe require competition authorities to define ``markets'' before assessing competitive effects, as highlighted in various legal cases and statutes. For example, the \replaced[id=PSK]{DOJ-FTC}{FTC-DOJ} Merger Guidelines have used both the level and the change in Herfindahl-Hirschman Index (HHI)---which depends on market definition---to determine whether a proposed merger should be presumed illegal.

Unfortunately, the use of market definition entails two critical complications. First, the term \emph{market definition} bears much ambiguity, and there is little consensus on its definition \citep{glasner2020logic}. A growing contingent of critics has argued for scrapping the market definition altogether and focusing solely on competitive effects analysis, which need not require arbitrary market delineation.\footnote{The hostility to market delineation \replaced[id=PSK]{has grown since the 1930s with the rise of monopolistic competition theory}{grew from the early 1930s with the development of the theory of monopolistic competition}, as product differentiation makes defining a relevant market ``an inevitably artificial line-drawing exercise'' \citep{werden1992history, farrell2010antitrust}. Louis Kaplow, in a series of papers \citep{kaplow2010ever, kaplow2011market, kaplow2012market, kaplow2013market, kaplow2015market, kaplow2021horizontal, kaplow2022replacing}, \replaced[id=PSK]{contends that market definition is redundant, producing unreliable market shares and relying on circular logic.}{argues that market definition serves no role except to produce market shares, which themselves serve as poor measures of market power, and that the logic of market delineation is inherently circular.} \replaced[id=PSK]{Despite its limitations, supporters highlight its practical and theoretical merits}{While acknowledging the limitations to market delineation, proponents argue that the practice of market delineation has practical merits on theoretical and empirical grounds} \citep{coate2012defense, werden2013ever, nocke2018multiproduct, glasner2020logic, miller2021quantitative, nocke2022concentration, nocke2023aggregative}.} However, the Supreme Court recently reaffirmed that the practice will not be abandoned soon: In \emph{Ohio v. American Express}\deleted[id=PSK]{585 US (2018)}, the Court disagreed with the plaintiffs' argument that market definition was not required because they had demonstrated adverse competitive effects\added[id=PSK]{: the Court explained that ``courts usually cannot properly apply the rule of reason without an accurate definition of the relevant market,'' thereby ruling that the plaintiffs failed to establish a \emph{prima facie} case \citep{supreme2018ohio}.} The case clarified \emph{when} market definition is needed but not \emph{how} it ought to be defined. 

Second, a single antitrust case may be subject to multiple market definitions. Multiple theories of harm may be relevant \added[id=PSK]{\citep{glasner2020logic}}. Moreover, the popular Hypothetical Monopolist Test may produce multiple market definitions \citep{baker2007market, davis2009quantitative, salop2009updating}. Perhaps most importantly, plaintiffs and defendants face uncertainty over which market definition will prevail in court. Thus,\added[id=PSK]{ for both plaintiffs and defendants,} scrutinizing the implications of alternative market delineation is a norm rather than an exception.\footnote{\added[id=PSK]{In the recent \emph{FTC v. Kroger} case, the FTC proposed two definitions as the relevant product market. First, it defined the ``Supermarkets'' market, which consists of traditional supermarkets and supercenters (e.g., Walmart and Target). Second, it also defined the ``Large-Format Stores'' market, which includes not only traditional supermarkets and supercenters but also club stores (e.g., Costco and Sam's Club), natural food stores (e.g., Whole Foods and Sprouts Farmers Market), and limited assortment stores (e.g., Trader Joe's and Aldi). The second market definition served to assess the sensitivity of the market concentration analysis to the market definition. The Court accepted both definitions \citep{FTC_v_Kroger_2024}.}} Despite the aforementioned debates and the pervasiveness of sensitivity analysis, attempts to establish a systematic approach have been lacking. This paper aims to fill this gap.

% MAIN RESULTS
In this paper, I propose a simple framework for conducting \replaced[id=PSK]{a form of}{an} \emph{organized sensitivity analysis}\added[id=PSK]{ \'{a} la \citet{leamer1985sensitivity}} with respect to market definition. The main contributions are twofold. First, I model the sensitivity analysis problem by leveraging the partial order nature of candidate markets and use \emph{Hasse diagram}---a directed acyclic graph for representing a finite partial order---to visualize the sensitivity of metrics of interest with respect to market definitions. My diagrammatic approach facilitates a compact summary of multi-dimensional measures from the sensitivity analysis. Second, I measure the importance of each firm (or a group of products/firms) using the \emph{Shapley value} and the \emph{Shapley-Shubik power index}. In its original form, the Shapley value is a solution concept from cooperative game theory that finds a fair allocation of gains among participants in a coalition by measuring each player's contribution; Shapley-Shubik power index is the Shapley value defined on simple games in which the outcome of a coalition is binary (0 or 1). I interpret the removal of firms from an antitrust market as the firms ``participating in a coalition'' to generate the final evaluation metric of interest (e.g., Herfindahl-Hirschman index or gross upward pricing pressure). Shapley value allows the analyst to understand the importance of each firm in determining the final conclusion.

% Application
I illustrate the framework's usefulness with an application to the Albertsons/Safeway merger consummated in 2015. Evaluating a supermarket merger requires drawing boundaries on which grocery chains should be considered sufficiently close competitors to traditional supermarkets. Many store formats serve grocery products to consumers but operate on heterogeneous business models. I examine the sensitivity of market concentration measures with respect to the inclusion and exclusion of wholesale club, natural/gourmet, and limited assortment stores. I also decompose the contribution of each major chain in triggering the structural presumption defined by the 2023 \replaced[id=PSK]{DOJ-FTC}{FTC-DOJ} Merger Guidelines\added[id=PSK]{ \citep{ftcdoj2023merger}}.

This paper contributes to the large literature on antitrust market definitions; see \citet{werden1992history}, \citet{baker2007market}, and \citet{davis2009quantitative} for reviews. While much effort has been dedicated to identifying the ``right'' market definition, the literature has paid less attention to sensitivity analysis\added[id=PSK]{, even though antitrust practitioners routinely rely on it to defend their arguments}. I fill this gap by developing novel quantitative tools for conducting organized sensitivity analysis with respect to market definition\added[id=PSK]{, which allows practitioners to formalize their sensitivity analysis problems and take a systematic approach}. \added[id=PSK]{I provide a novel application of Hasse diagrams for sensitivity analysis, which adds to a set of diagrammatic tools economists can use to summarize complex information (e.g., \citet{pearl1995causal}'s causal diagram).} \added[id=PSK]{Furthermore, }I find a novel application of Shapley value, a well-known cooperative game theory solution concept primarily applied to study voting power. The use of Shapley value for sensitivity analysis has been increasing in the statistical and machine learning literature \citep{owen2014sobol, song2016shapley, plischke2021computing, rozemberczki2022shapley}. To the best of my knowledge, quantitative sensitivity analysis with respect to market definitions is new.

My sensitivity analysis framework also complements empirical tools that hinge on market definitions. While I focus on HHI as an evaluation metric throughout the paper for simplicity\added[id=PSK]{ (and its popularity)}, my framework extends to various statistics that may be of interest, including concentration ratio, Herfindahl-Hirschman index \citep{hirschman1964paternity, nocke2018multiproduct, nocke2022concentration, nocke2023aggregative}, merger simulation \citep{nevo2000mergers}, compensating marginal cost reduction \citep{werden1996robust}, upward pricing pressure \citep{farrell2010antitrust, conlon2021empirical, miller2021quantitative, koh2024merger}.\footnote{\replaced[id=PSK]{Upward pricing pressure and compensating marginal cost reductions can bypass market definition with reliable diversion ratio data. However, they often rely on it, as diversion ratios are estimated using market-based formulas (e.g., logit or CES) or demand models sensitive to market size assumptions and consumer consideration sets }{In principle, upward pricing pressure and compensating marginal cost reductions need not rely on market definition (if reliable diversion ratio data are available). On many occasions, however, they do because diversion ratios need to be estimated. First, especially in the merger screening stage, it is common for economists to specify the demand function as logit so that diversion ratios can be approximated with the proportional-to-share formula, which depends directly on market definition. Second, even flexible demand models (e.g., random-coefficients logit models) can be sensitive to market size parameters or the set of products included in consumers’ consideration set, which relate to market definition}\citep{nevo2000practitioner}. } More broadly, the framework may be applied to other settings where the analyst wants to report the sensitivity of the main results with respect to a partially ordered discrete set of assumptions.

Note that this paper does not attempt to select the right market definition nor respond to the current debate over market definition. This paper also does not seek to develop a refinement over the current economic tools for analyzing mergers. Instead, I take the current legal standards around market definition as given and propose a framework for sensitivity analysis---which is \emph{always} carried out---based on classical game-theoretic tools. My framework can help aid robustness analysis and guide resource allocation.

%%% Outline of the paper %%%
The rest of the paper is organized as follows. Section \ref{section:model.of.sensitivity.analysis} presents a model of sensitivity analysis and the use of Hasse diagrams. Section \ref{section:measuring.the.importance} illustrates how the Shapley value can measure the importance of each firm in market definition. Section \ref{section:empirical.application} considers an empirical application. Section \ref{section:conclusion} concludes. \added[id=PSK]{The replication package is available at \url{https://doi.org/10.5281/zenodo.16809457}.}

% To the best of my knowledge, \citet{cheung2016antitrust} is the only paper that studies the sensitivity with respect to market definition.

\section{A Model of Sensitivity Analysis \label{section:model.of.sensitivity.analysis}}
Expanding the product and/or geographic markets adds more competitors and reduces each firm's market share. Many popular statistics for evaluating market power and competitive effects of mergers---e.g., concentration ratio, Herfindahl-Hirschman index (both in absolute level and changes from pre-merger level to post-merger level), compensating marginal cost reduction \citep{werden1996robust}, and upward pricing pressure \citep{farrell2010antitrust}---are functions of market shares, and expanding market definition generally leads to more conservative (lower) estimates. For ease of exposition, I focus on a scenario in which the analyst is interested in examining a model's sensitivity to the \emph{exclusion} of potential competitors so that expanding the list of firms to be removed from the market \emph{increases} the evaluation metric.

Let $\mathcal{N} \equiv \{1,2,...,N\}$ be the set of marginal firms under consideration. Each $\omega \in 2^\mathcal{N}$ represents an \emph{exclusion set} defined as the list of firms to be dropped from the market. Function $f: 2^\mathcal{N} \to \mathbb{R}$, which I refer to \added[id=PSK]{as }the \emph{outcome function}, returns the value of an evaluation metric. The analyst's objective is to study the sensitivity of $f$ with respect to changing $\omega$. I provide a running example below.

\begin{example}\label{example:1}
    Suppose the universe of all firms is $\{A, B, 1, 2, 3\}$. Firms A and B announce a merger. To evaluate its impact on market concentration, the analyst wants to compute HHI but is unsure whether Firms 1, 2, or 3 should enter the relevant market definition. In this case, the set of marginal members under consideration is $\mathcal{N} = \{1,2,3\}$. The function $f(\omega)$ measures the HHI excluding firms in $\omega \in 2^{\{1,2,3\}}$ from $\{A, B, 1, 2, 3\}$. For example, $f(\{1,3\})$ measures the HHI after taking the relevant market to be $\{A,B,2\}$. 
\end{example}

% \begin{assumption}\label{assumption:model.primitives}
% I specify the primitives of the model by a tuple
% \begin{equation}\label{equation:model.primitives}
% \langle (\Omega, \subseteq), (\mathbb{R}, \leq), f \rangle.
% \end{equation}
% Set $\Omega = \mathcal{P}(\mathcal{J})$ is the power set defined over a set of firms $\mathcal{J} = \{1,...,J\}$, and $\subseteq$ is the inclusion relation. Function $f:\Omega \to \mathbb{R}$ is the measure of outcome; given any subset of firms $\omega \in \mathcal{P}(\mathcal{J})$, $f(\omega)$ is a measure the analyst use to base decisions.     
% \end{assumption}

\subsection{Hasse Diagram Representation of Finite Partial Order}
The partial order $(2^\mathcal{N}, \subseteq)$, where $\subseteq$ represents the inclusion relation\footnote{\added[id=PSK]{An inclusion relation is defined as $A \subseteq B$ if and only if $A$ is a subset of $B$.}}, formalizes the idea that a market definition may be broader or narrower than another, but two markets need not be comparable. To graphically represent the partial order, I use a Hasse diagram, a mathematical diagram representing a transitive reduction ordering of a finite partially ordered set. 
\begin{definition}[Hasse diagram]\label{definition:hasse.diagram}
Let $(S, \precsim)$ be a finite partial order.\footnote{\added[id=PSK]{A finite partial order is a pair of finite set $S$ and a partial order $\precsim$. A partial order is a binary relation that satisfies reflexivity (for all $x \in S$, $x \precsim x$), antisymmetry (for all $x,y \in S$, if $x \precsim y$ and $y \precsim x$, then $x = y$), and transitivity (for all $x,y,z\in S$, if $x \precsim y$ and $y \precsim z$, then $x \precsim z$).}} A \emph{Hasse diagram} of $(S, \precsim)$ is a graph $G$ in which (i) the vertices of $G$ represent the elements of $S$; (ii) the edges of $G$ represent the elements of $\precsim$; (iii) if $x,y \in S$ and $x \prec y$, then the point corresponding to $x$ appears lower in the drawing than the point corresponding to $y$; (iv) the line segment between the points corresponding to any two elements $x$ and $y$ of the partially ordered set is included in the drawing if and only if $x$ covers $y$ or $y$ covers $x$.\footnote{\replaced[id=PSK]{Formally, element}{Element} $x$ covers $y$ if $x \neq y$ and $y \precsim x$ and there does not exist $z$ such that $y \precsim z \precsim x$.}
\end{definition}

Any finite partial order admits a Hasse diagram representation; however, a Hasse diagram representation for a given finite partial order may not be unique. I provide an example of a Hasse diagram as follows.

\begin{figure}[htbp!]
    \centering
    \includegraphics[scale=0.65]{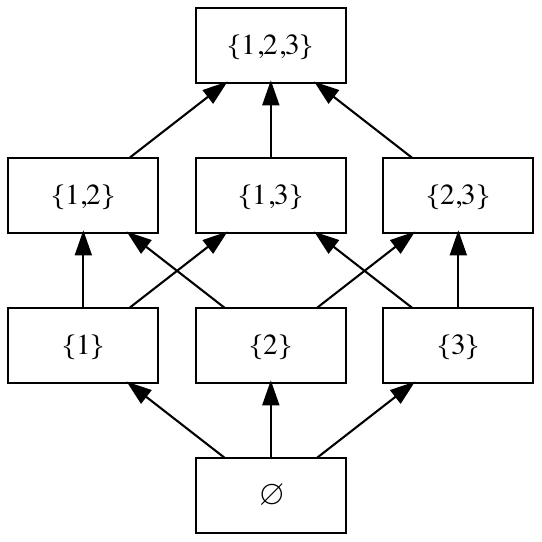}
    \caption{An Example of Hasse Diagram}
    \label{figure:hasse.diagram.example}
\end{figure}

\setcounter{example}{0}
\begin{example}[Continued]
Figure \ref{figure:hasse.diagram.example} is a Hasse diagram for $(2^\mathcal{N}, \subseteq)$ where $\mathcal{N} = \{1,2,3\}$. With nodes defined by the elements of $2^\mathcal{N}$ and edges defined by the inclusion relation, the directed acyclic diagram compactly summarizes the hierarchical relationships among the subsets of $\{1, 2, 3\}$. The bottom node $\emptyset$ represents the broadest market where no firm is excluded. The top node $\{1,2,3\}$ corresponds to the narrowest market in which all firms in $\{1,2,3\}$ are excluded, and only the merging firms $\{A,B\}$ remain. Note that edges are drawn only between nodes that have a covering relationship; for example, there is no direct arrow from $\{1\}$ to $\{1,2,3\}$ since $\{1,2\}$ (or $\{1,3\}$) is positioned between them. 
\end{example}

It is always possible to draw a Hasse diagram with a subset of nodes since any subset of nodes form a partial order, and any finite partial order admits a Hasse diagram representation. Drawing a Hasse diagram with a subset of nodes can be helpful when the power set of $\mathcal{N}$ is too large and when particular subsets of firms are of primary interest.

\subsection{Measures of Competitive Effects}
It is possible to augment the diagram by layering each node with the value of the outcome function. In a typical application, $f:2^\mathcal{N} \to \mathbb{R}$ is monotonic. In \ref{section:monotonicity}, I establish the monotonicity of popular statistics (concentration ratios, HHI, upward pricing pressure, and compensating marginal cost reduction) with respect to market definition.\footnote{I emphasize again that while upward pricing pressure and compensating marginal cost do not rely on market definition (if reliable diversion ratio data are available), practitioners often rely on logit\added[id=PSK]{ (or CES)} demand assumptions to approximate diversion ratios using cross-sectional market shares data.}

\setcounter{example}{0}
\begin{example}[Continued]
Suppose that there are a total of eight firms with sales $x_A = 15$, $x_B = 15$, $x_C = 10$, $x_D = 10$, $x_E = 10$, $x_1 = 9$, $x_2 = 6$, $x_3 = 3$. I assume firms $\{A,B,C,D,E\}$ are always included in the market but test the sensitivity of HHI with respect to the exclusion of Firms 1, 2, and 3. Figure \ref{figure:hasse.diagram.example.2} summarizes the sensitivity of pre-merger HHI with $\mathcal{N} = \{1,2,3\}$.\footnote{\added[id=PSK]{The values are rounded to the nearest integer.}} Each node reports the level of HHI for a given market definition, and each edge reports the marginal change in HHI as the market definition narrows. For example, the HHI associated with the broadest market definition (bottom node) is 1439. However, excluding Firm 1 increases the HHI by 230 to 1669. Excluding Firms 1, 2, and 3 leads to an HHI of 2083. In general, as shown in \ref{section:monotonicity}, excluding more firms in the relevant market increases HHI insofar as the excluded firms' sales are not too large. 

\begin{figure}[htbp!]
    \centering
    \includegraphics[scale=0.70]{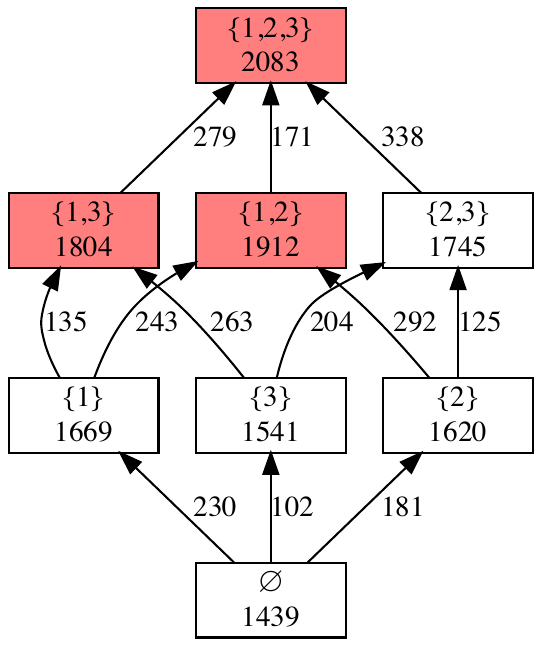}
    \caption{Annotated Hasse Diagram}
    \label{figure:hasse.diagram.example.2}
\end{figure}    

I assume that the HHI threshold for a highly concentrated market is 1,800, as defined in the 2023 FTC-DOJ Merger Guidelines. In Figure \ref{figure:hasse.diagram.example.2}, the nodes corresponding to high-HHI markets are shaded in red. It shows that excluding Firm 1 is essential for triggering the high-HHI threshold, but either Firm 2 or 3 needs to be excluded from the market as well.
\end{example}

As seen from the example above, Hasse diagrams can organize complex information based on the partial order of market definitions. Although I have assumed a single outcome function $f$, it is possible to augment the diagram with multidimensional information. Yet, the complexity of the diagram quickly rises \replaced[id=PSK]{with}{as} the number of firms in $\mathcal{N}$. Specifically, the Hasse diagram for $(2^\mathcal{N}, \subseteq)$ has $2^N$ nodes and $N \cdot 2^{N-1}$ edges. In the following section, I propose using the Shapley value to measure the average contribution of each firm.

\section{Sensitivity Analysis with Shapley Value \label{section:measuring.the.importance}}
Suppose that the outcome function $f$ is monotonic such that the broadest market definition generates the most conservative (lowest) estimate of market power and vice versa. Then how much does each firm contribute in driving the output from $f(\emptyset)$ to $f(\mathcal{N})$? I show that the Shapley value provides a simple answer to this question. 

\subsection{Shapley Value for Sensitivity Analysis}
I begin by reviewing the definition of coalitional games and Shapley values as follows.
\begin{definition}[Shapley value] \label{definition:shapley.value}
A pair $\langle \mathcal{N}, v \rangle$ is a \emph{coalitional game} if $\mathcal{N}$ is a finite set of players and $v:2^\mathcal{N} \to \mathbb{R}$ is a characteristic function such that $v(\emptyset) = 0$. The \emph{Shapley value} for each player $i$ in a coalitional game $\langle \mathcal{N}, v \rangle$ is
\begin{equation}\label{equation:shapley.value}
    \varphi_{i} = \sum_{S \subseteq N \backslash \{i\}} \frac{\vert S \vert ! (N- \vert S \vert - 1)!}{N!}(v(S \cup \{i\})- v(S)).
\end{equation}
\end{definition}
Shapley value \citep{shapley1953value} is a single-valued solution concept for coalition games that proposes how to distribute the total gains to the players, assuming that they all collaborate.\footnote{Shapley value is defined by an axiomatic approach: it is the unique solution concept that satisfies the efficiency, symmetry, null player, and additive properties. See \citet*{maschler2020game} for a textbook treatment of the topic. It is possible to extend the definition and assign asymmetric weights to players \citep{shapley1953additive}. } In Definition \ref{definition:shapley.value}, the function $v(S)$ describes the value of coalition $S \in 2^\mathcal{N}$. For each $S \in 2^\mathcal{N}$ such that $i \notin S$, $v(S \cup \{i\}) - v(S)$ is the marginal surplus that player $i$ brings to the table. The Shapley value of $i$ is the average marginal contribution over all possible $S$ that permits such calculation. Thus, \eqref{equation:shapley.value} measures the average marginal contribution of each player over the possible different permutations in which the coalition can be formed. A key property of Shapley value is that
\begin{equation}\label{equation:shapley.value.sum}
    \sum_{i=1}^N \varphi_i = v(\mathcal{N}),
\end{equation}
i.e., the sum of Shapley values is equal to the total value of the grand coalition. \deleted[id=PSK]{Shapley value measures how much the players can expect from a fair distribution of total surplus generated by the grand coalition.}

Shapley value has a natural application to sensitivity analysis because it decomposes the contribution of each member in generating the final output. I interpret the removal of firms in the relevant market as the firms forming a coalition. For example, if firms 1 and 2 are removed from the market definition, it is as if they formed a coalition to exclude themselves from the market. Since the set of firms $\mathcal{N} =\{1,...,N\}$ is given, the description of a coalitional game is completed by defining a characteristic function. For each subset of firms $S \in 2^\mathcal{N}$, define the characteristic function as
\begin{equation}\label{equation:characteristic.function}
    v(S) \equiv f(S) - f(\emptyset).
\end{equation}
The characteristic function \eqref{equation:characteristic.function} is simply an outcome function normalized so that $v(\emptyset) = 0$. The Shapley value measures the importance of each firm in increasing the outcome measure by $v(\mathcal{N})$ when going from the broadest market definition to the narrowest market definition.

\setcounter{example}{0}
\begin{example}[Continued]
Recall from Figure \ref{figure:hasse.diagram.example.2} that $f(\emptyset) = 1439$ and $f(\{1,2,3\}) = 2083$. Thus, excluding firms $\{1,2,3\}$ from the market increases the level of HHI by $v(\{1,2,3\}) \equiv f(\{1,2,3\}) - f(\emptyset) = 2083 - 1439 = 644$. Table \ref{table:shapley.value.example} reports the Shapley values of each firm.\footnote{I use \texttt{R} package \texttt{CoopGame} \citep{staudacher2019coopgame} to compute the Shapley values.}\added[id=PSK]{ To calculate the Shapley value of Firm 1, for instance, note that there are $3!=6$ ways to sequentially form the grand coalition: 1-2-3, 1-3-2, 2-1-3, 3-1-2, 2-3-1, 3-2-1; each sequence represents a feasible path from node $\emptyset$ to node $\{1,2,3\}$ in the Hasse diagram. In the first two cases, Firm 1 is the first to enter the coalition, so the marginal contribution of Firm 1 is $\nu(\{1\})-\nu(\{0\})=229.9751$. In the third and fourth cases, the marginal contribution of Firm 1 is $\nu(\{1,2\})-\nu(\{2\}) = 291.9501$ and $\nu(\{1,3\}) - \nu(\{3\}) = 263.0744$, respectively. In the last two cases, Firm 1 is the last to enter the coalition, so it is $\nu(\{1,2,3\})-\nu(\{2,3\})=337.9017$. Taking the average over the six permutations gives $281.7693 \approx 282$.}  Note that the Shapley values sum to the value of the grand coalition 644.

\begin{table}[htbp!]
    \centering
    \caption{Shapley Value and Shapley Shubik Power Index \label{table:shapley.value.example}}
    \begin{threeparttable}
    \begin{tabular}{cccccccc} \toprule
    Firm & Sales & Sales Share & & SV & SV Share & & SSPI \\ \midrule
    1 & 9 & 50\% & & 282 & 44\% & & 0.666 \\ 
    2 & 6 & 33\% & & 228 & 35\% & & 0.166 \\ 
    3 & 3 & 17\% & & 134 & 21\% & & 0.166 \\ \midrule
    Total & 60 & 100\% & & 644 & 100\% & & 1.0 \\ \bottomrule
    \end{tabular}
    \begin{tablenotes}
        \footnotesize
        \item \added[id=PSK]{\emph{Notes:} Shapley values (SV) are rounded to the nearest integer.}
    \end{tablenotes}
    \end{threeparttable}
\end{table}

Column ``SV Share'' says that Firms 1, 2, and 3 account for 44\%, 35\%, and 21\% of the total increment. The share of average contribution measured by Shapley value differs from the sales shares, which are 50\%, 33\%, and 17\% for Firms 1, 2, and 3, respectively. Using the sales shares to explain the importance of each firm in the HHI calculation overstates Firm 1's contribution and understates Firm 3's contribution. 
\end{example}

Table \ref{table:shapley.value.example} is significant because it breaks down each firm's contribution to the final outcome measure of interest. In this example, while the sales shares and Shapley values are roughly proportional, using Shapley values offers a distinct advantage. The analyst derives the Shapley values directly from the outcome values, which depend nonlinearly on shares, whereas sales shares merely preserve rankings. In the following, I demonstrate that the utility of Shapley values is even greater when the outcome measure represents a binary decision rule, which is ultimately what matters.

% \begin{remark}
%     Harsanyi dividend \citep{harsanyi1959bargaining} identifies the surplus created by each coalition of players in a coalition game. It offers a unifying framework for studying various valuation concepts. The dividend of coalition $S$ is computed by $d(S) = \sum_{T \subseteq S} (-1)^{\vert S \backslash T\vert} v(T)$. Harsanyi dividend can be useful for analyzing games and studying various solution concepts. The Shapley value is obtained as $\phi_i = \sum_{S \subseteq N: i \in S} d(S) / \vert S \vert$.
% \end{remark}

\subsection{Shapley-Shubik Power Index}
I show that Shapley-Shubik Power Index \citep{shapley1954method} can quantify the relative importance of each member in triggering a binary decision such as the structural presumption for merger screening. 

\begin{definition}
A coalitional game in which the worth of each coalition is 0 or 1 is called a \emph{simple game}, i.e., the outcome function is $f: 2^\mathcal{N} \to \{0,1\}$. When applied to simple games, the Shapley value is known as the \emph{Shapley-Shubik power index}.
\end{definition}
Intuitively, the Shapley-Shubik power index of a player is the probability that the player will be a pivot player; in this sense, the index measures the power of each player in turning the outcome from 0 (``loss'') to 1 (``win'').\added[id=PSK]{ Shapley-Shubik power index is also calculated using formula \eqref{equation:shapley.value}, but with a characteristic function that has range $\{0,1\}$.} Shapley-Shubik power index is widely used in political science as a measure of power distribution in committees.

Many antitrust applications can be cast as simple games because certain binary decisions are triggered depending on how the relevant market is defined. For example, the 2023 \replaced[id=PSK]{DOJ-FTC}{FTC-DOJ} Merger Guidelines presume that a merger substantially lessens competition or tends to create a monopoly if either (i) the post-merger HHI is greater than 1,800, and the merger increases the HHI by more than 100, or (ii) the merged firm's market share is greater than 30\%, and the merger increases the HHI by more than 100. The analyst can encode the outcome as 1 if a given market definition triggers the structural presumption and 0 otherwise. Removing more competitors will likely increase the chance of triggering the structural presumption. The Shapley-Shubik power index measures the importance of each member in triggering the presumption.

\setcounter{example}{0}
\begin{example}[Continued]
Suppose the analyst wants to determine whether the pre-merger market is highly concentrated. A market is deemed highly concentrated if the HHI is higher than 1800. Recall that $v(S)$ measures the pre-merger HHI associated with coalition $S$. The decision rule induces a simple game with characteristic function
\begin{equation}
    v^*(S) = 
    \begin{cases}
        1 & \quad \text{if} \quad v(S) \geq 1800, \\
        0 & \quad \text{if} \quad v(S) < 1800.
    \end{cases}
\end{equation}
\added[id=PSK]{The Shapley-Shubik power indices can be calculated following the same steps as in the case of Shapley value, but with characteristic function $\nu^*$. For example, consider Firm 1. Firm 1's marginal contribution for each sequence is calculated as follows. For sequences 1-2-3 and 1-3-2, Firm 1's marginal contribution is $\nu^*(\{1\})-\nu^*(\emptyset)=0$; for sequence 2-1-3, $\nu^*(\{1,2\})-\nu^*(\{2\})=1$; for sequence 3-1-2, $\nu^*(\{1,3\})-\nu^*(\{3\})=1$; finally, for sequences 2-3-1 and 3-2-1, $\nu^*(\{1,2,3\})-\nu^*(\{2,3\})=1$. Taking the average over the 6 possible sequences of coalition formation gives $2/3\approx 0.67$. }

Column ``SSPI'' of Table \ref{table:shapley.value.example} reports the Shapley-Shubik power index derived from $(\mathcal{N}, v^*)$. First, it gives Firm 1 a weight of 67\%, far higher than the weights of the other two firms. Second, surprisingly, Firms 2 and 3 have the same power index values, indicating that the two firms have equal importance in triggering the decision even though one firm has twice the market share as the other. Thus, the example illustrates why it is incorrect to assume that the importance of each firm in triggering a decision is strictly proportional to market shares.
\end{example}

\subsection{Discussion}

\subsubsection*{Choice of Outcome Function}
\added[id=PSK]{The choice of an outcome function $f:2^\mathcal{N} \to \mathbb{R}$ determines (i) the outcome measure of interest (e.g., market concentration or whether the structural presumption is triggered), and (ii) the dimension along which the analyst seeks to check the sensitivity of the outcome measure (e.g., which products should enter the antitrust market). Whether the researcher obtains Shapley value or the Shapley-Shubik power index merely depends on the support of $f$ since Shapley-Shubik power index is only a Shapley value applied to cooperative games with binary outcomes (which are also called ``voting games'' in the cooperative game literature). The choice of outcome function $f$ is quite flexible but requires the domain to be of form $2^\mathcal{N}$ and the codomain to be a real space.}\footnote{\added[id=PSK]{Note, however, a Hasse diagram can visualize any partial order. The requirement that the domain is $2^\mathcal{N}$ is only relevant for Shapley value calculations.}} \added[id=PSK]{Thus, the applicability of my framework crucially depends on whether the analyst can cast the sensitivity analysis exercise as the outcome being a function of a vector of binary decisions.}

\subsubsection*{Choice of Sensitivity Measure}
\added[id=PSK]{Although I have focused on the standard Shapley value (and Shapley-Shubik power index, which is a special case) as a tool for quantifying the importance of each input component, the researcher may want to apply a different approach. For example, a possible approach is to consider a weighted Shapley value where asymmetric weights are assigned to the participants of a cooperative game \citep{shapley1953additive}. In antitrust applications, the analyst might want to assign asymmetric weights to products based on prior knowledge and documentary evidence. However, Shapley value provides a natural starting point as the concept is easy to calculate and interpret. Shapley value is also the only allocation rule that satisfies four properties (efficiency, symmetry, additivity, and the null player property) deemed desirable for an allocation rule \citep{shapley1953value}. Modifying the formula (e.g., assigning asymmetric weights to players) may lead to a measure that violates these properties.}\footnote{\added[id=PSK]{See \citet{maschler2020game} for examples of alternative distribution rules that violate the Shapley properties.}} 

\subsubsection*{The Hypothetical Monopolist Test}

\added[id=PSK]{My framework can complement the Hypothetical Monopolist Test (HMT), a tool often used in antitrust analysis to define the relevant market in merger or competition cases.}\footnote{\added[id=PSK]{I thank an anonymous referee for making this suggestion.}} \added[id=PSK]{For example, consider an outcome function that shows how a hypothetical monopolist changes price (obtained as the solution to the hypothetical monopolist's profit maximization problem) relative to pre-merger values at different candidate markets. The analyst can use an analog of Figure \ref{figure:hasse.diagram.example.2} to report the sensitivity of predicted price changes with respect to the inclusion or exclusion of marginal products. Such an exercise can inform the analyst whether the outcome of the HMT is sensitive to the choice of SSNIP threshold, especially given that the typical choice of 5\% as the SSNIP threshold is often criticized as arbitrary. Shapley values calculated following the steps in Example \ref{example:1} can inform the relative importance of each marginal product in determining the final antitrust market.}

\added[id=PSK]{I note that although the SSNIP test is a popular approach to finding relevant antitrust markets, implementing it can be difficult in practice because it requires the researcher to have a model that predicts how consumers respond to price changes, which may require demand elasticity estimates, potentially from complex demand models (e.g., \citet{berry1995automobile}). For this reason, practitioners often rely on other metrics such as HHI, $\Delta$HHI, and UPP, all of which serve as useful complements for examining issues related to market definitions and can be embedded in my framework.}

\section{Empirical Application to Albertsons/Safeway (2015) \label{section:empirical.application}}

In this section, I apply the proposed framework to the Albertsons/Safeway merger. I analyze the sensitivity of the HHI calculation and structural presumption with respect to the exclusion of club, natural, and limited assortment stores, which may or may not be considered close competitors to traditional supermarkets.\footnote{I make no claims on which formats should be included in the relevant market definition when evaluating supermarket mergers.} I consider both state-level and local-level analyses (see below for the definition of local markets).

\subsection{Background}
In March 2014, the Albertsons supermarket chain entered into an agreement to purchase Safeway for \$9 billion. Albertsons operated 1,075 stores in 28 states under the banners Albertsons, United, Amigos, and Market Street, among others; Safeway owned 1,332 stores in 18 stores under the banners Safeway, Vons, Pavilions, Tom Thumb, and Randall's, among others. If consummated, the merger was to create the second-largest traditional grocery chain by store count and sales in the US. In late 2014, the FTC settled with the parties with a mandate to divest stores and production facilities.

The FTC defined the relevant product market as supermarkets within ``hypermarkets'' \added[id=PSK]{\citep{ftc2015analysis}}. Supermarkets refer to ``traditional full-line retail grocery stores that sell, on a large-scale basis, food and non-food products that customers regularly consume at home---including, but not limited to, fresh meat, dairy products, frozen foods, beverages, bakery goods, dry groceries, detergents, and health and beauty products.'' Hypermarkets include chains such as Walmart Supercenters that sell an array of products not found in traditional supermarkets but also offer goods and services available at conventional supermarkets. The relevant geographic markets were defined as areas that range from a two- to ten-mile radius around each of the defendants' stores, where the relevant radius varies depending on factors such as population density, traffic patterns, and unique characteristics of each market.

Historically, defining the relevant antitrust market for supermarkets has been challenging for several reasons. Supermarkets offer a wide range of products beyond groceries, including household items, pharmacy goods, and electronics. Consumers also shop at multiple types of stores for groceries, including wholesale clubs (e.g., Costco, Sam's Club), limited assortment stores (e.g., Aldi, Lidl), natural/gourmet stores (e.g., Whole Foods, Sprouts), and ethnic stores (e.g., H Mart, Fiesta Mart). Traditionally, the FTC defined the relevant market as ``full-line grocery stores'' that provide ``one-stop shopping'' that enables consumers to ``shop in a single store for all of their food and grocery needs.''\footnote{See, for example, Albertsons/Safeway, Ahold/Delhaize, \added[id=PSK]{Kroger/Albertsons}, and Whole Foods/Wild Oats mergers.} The debate over market definition was particularly intense during Whole Foods/Wild Oats: While the FTC contended that the relevant market should be defined as ``premium, natural, and organic supermarkets,'' the merging parties asserted that the relevant product market should include all supermarkets.

\subsection{Data}
I illustrate how the proposed approaches can facilitate sensitivity analysis with respect to market definition. I study grocery chains' market shares in the State of Washington, where Albertsons and Safeway had a significant overlap in their footprint. I use the 2009 cross-section of the AC Nielsen's (currently known as The Nielsen Company) Trade Dimensions data, which reports retail stores' locations, revenues, and characteristics.\footnote{Year 2009 is the closest year before 2014 for which I have access to the Trade Dimensions data. To mimic the Albertsons/Safeway merger investigation that started in 2014, I encode Albertsons stores in Washington State as owned by Albertsons; SuperValu Inc. had ownership over the Albertsons banner in the region but sold it to Cerberus Capital Management (which owned the Albertsons company) in 2013.} The Trade Dimensions dataset categorizes each grocery store into one of multiple formats. I categorize stores into ``supermarket,'' ``supercenter,'' ``club,'' ``natural,'' and ``limited.''\footnote{Specifically, I categorize a store as a ``supermarket'' if it is categorized as a supermarket or a superette (small supermarkets) in the Trade Dimensions data; ``supercenter'' if categorized as a supercenter; ``club'' if categorized as a wholesale club or a warehouse; ``natural'' if categorized as a natural/gourmet store; ``limited'' if categorized as a limited assortment tore.} I include all grocery stores as potential competitors but drop military commissary stores.

\begin{table}[htbp!]
\centering
\begin{threeparttable}
\caption{Summary Statistics \label{table:summary.statistics}}
\begin{tabular}{lccccc} \toprule
Variable & Mean & SD & Min & Median & Max \\ \midrule
Revenue (\$mil) & 17.33 & 16.43 & 2.03 & 13.04 & 95.16 \\
Supermarket & 0.74 & 0.44 & 0.00 & 1.00 & 1.00 \\
Supercenter & 0.09 & 0.29 & 0.00 & 0.00 & 1.00 \\
Club & 0.07 & 0.25 & 0.00 & 0.00 & 1.00  \\
Natural/Gourmet & 0.06 & 0.24 & 0.00 & 0.00 & 1.00 \\
Limited Assortment & 0.04 & 0.20 & 0.00 & 0.00 & 1.00 \\ \midrule 
\added[id=PSK]{Number of Stores} & \added[id=PSK]{859} \\ \bottomrule
\end{tabular}    
\begin{tablenotes}
    \footnotesize
    \item \emph{Notes}: The sample consists of all grocery stores in Washington State (excluding military commissary stores).
\end{tablenotes}
\end{threeparttable}
\end{table}

Table \ref{table:summary.statistics} reports the basic summary statistics. In 2009, \added[id=PSK]{there were 859 grocery stores in Washington State.}\footnote{\added[id=PSK]{The dataset contains 864 stores, but I exclude five that have zero revenue.}} An average grocery store earned \$17.23 million in revenue, but the revenues could vary substantially across stores. The dominant formats (in \added[id=PSK]{terms of} store counts) were traditional supermarkets and supercenters, which accounted for 74\% and 10\% of all stores, respectively. Club, natural, and limited assortment stores accounted for less than 20\% of all Washington State stores.

\begin{table}[htbp!]
\centering
\caption{Top 11 Grocery Chains in Washington State \label{table:top.9.chains}}
\begin{tabular}{l c *{3}{S[table-format=4.3]}} \toprule
Firm & Primary Format & $\text{Revenue (\$mil)}$  & $\text{Market Share}$  & \text{Store Count}  \\ \midrule
Kroger & Supermarket & 3239.37 & 0.217 & 128 \\
Safeway & Supermarket & 3058.17 & 0.205 & 168 \\
Costco & Club & 1862.80 & 0.125 & 29 \\
Walmart & Supercenter & 1805.44 & 0.121 & 34 \\
Albertsons & Supermarket & 1062.41 & 0.071 & 75 \\
Winco & Supermarket &  508.39 & 0.034 & 12 \\
Haggen & Supermarket &  449.73 & 0.030 & 29 \\
Trader Joes & Natural & 241.16 & 0.016 & 15 \\
The Markets, LLC & Supermarket & 195.54 & 0.013 & 21 \\
Whole Foods & Natural & 181.19  & 0.012 & 5 \\ 
Grocery Outlet & Limited & 170.71 & 0.011 & 30 \\
 \bottomrule
\end{tabular}
\end{table}

Table \ref{table:top.9.chains} reports the revenue from grocery sales, market share, and the number of stores for the top 11 grocery chains in Washington State. The market shares in the table are calculated using all grocery stores in the sample. The top five grocery chains account for almost three-quarters of the total revenue in the sample. Costco and Walmart combined had a significant presence with approximately 25\% market share in revenue. Trader Joe's and Whole Foods were the largest natural grocery retail chains. Grocery Outlet was the largest limited assortment chain. \added[id=PSK]{Note that Safeway and Albertsons have 243 stores in WA combined.}

\subsection{State-Level Analysis}
As a first exercise, I ignore the spatial nature of grocery competition and calculate \added[id=PSK]{grocery} chains' market shares at the state level. The primary outcome measure of interest is whether the merger triggers structural presumptions. In this exercise, following the 2023 \replaced[id=PSK]{DOJ-FTC}{FTC-DOJ} Merger Guidelines, I assume that a merger is presumed illegal if the post-merger HHI is above 1,800 and the HHI increases by more than 100. I examine the sensitivity of the outcomes with respect to grocery store formats. I always include supermarkets and supercenters in the market definition. I test how the outcomes change when I exclude club stores, limited assortment stores, or natural stores. That is, I set $\mathcal{N} = \{\text{Club}, \text{Natural}, \text{Limited} \}$.  \added[id=PSK]{Since I only observe store-level revenues, I use revenue-based HHIs.}\footnote{\added[id=PSK]{Both revenues and quantities can be used to calculate HHI.}}

\begin{figure}[htbp!]
    \centering
    \includegraphics[scale=0.60]{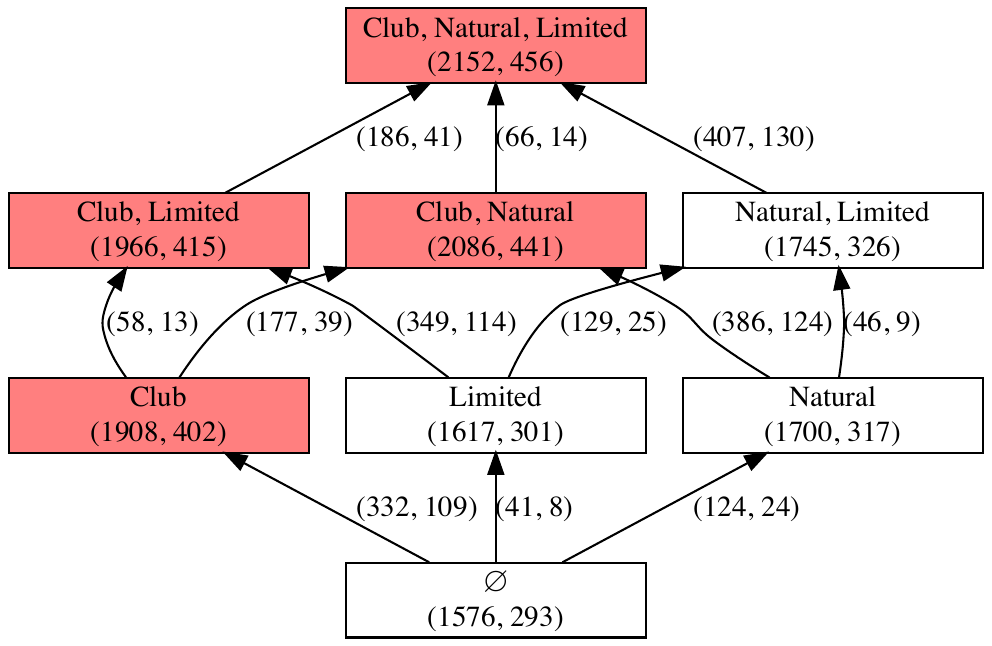}
    \caption{Post-Merger HHI and $\Delta$HHI} \label{figure:albertsons.safeway.hasse.diagram}
\end{figure}

Figure \ref{figure:albertsons.safeway.hasse.diagram} illustrates how the post-merger HHI and $\Delta$HHI (defined as the difference between post-merger HHI and pre-merger HHI) vary as I exclude more formats from the market. Each pair in the nodes represents the values of post-merger HHI and $\Delta$HHI under the assumed market definition. Nodes shaded in red correspond to market definitions that trigger the structural presumption. The values in the edges represent the changes in post-merger HHI and $\Delta$HHI when shrinking the market definition by excluding certain formats.

When I include all grocery formats (supermarkets, supercenters, clubs, limited assortment stores, and natural stores), the Albertsons/Safeway merger produces the post-merger HHI of 1,576 and $\Delta$HHI of 293 in Washington State.\footnote{Post-merger HHI of 1,576 falls short of being labeled as ``highly concentrated'' in the 2023 Merger Guidelines. However, the 1982 version classified HHI between 1,000 and 1,800 as ``moderately concentrated'' until 2010.}  Excluding club stores triggers the structural presumption. However, excluding natural or limited assortment stores from the market definition does not affect the structural presumption, as doing so only slightly increases the post-merger HHI. Removing all club, natural, and limited stores raises the post-merger HHI and $\Delta$HHI to 2,152 and 456, respectively.

Table \ref{table:empirical.application.shapley.value} reports the post-merger HHI Shapley value of each format. Out of the total increment of 576 ($=2152 - 1576$) in the post-merger HHI resulting from removing the three formats, club, natural, and limited assortment formats account for 64.0\%, 26.8\%, and 9.2\%, respectively. Thus, club format alone accounts for more than half of the increase in post-merger HHI, even though the number of club stores is small relative to other formats. The SSPI column, which reports the Shapley-Shubik power index of each format, reveals that club format is a ``dictator'': triggering the structural presumption only depends on whether club stores are included; natural and limited assortment formats have no role. 

\begin{table}[htbp!]
\centering
\caption{Shapley Values for Post-merger HHI and Structural Presumption \label{table:empirical.application.shapley.value}}
\begin{tabular}{lccc} \toprule
Format & SV & SV Share & SSPI \\ \midrule
Club & 386.71 & 0.640 & 1 \\
Natural & 154.23 & 0.268 & 0 \\
Limited & 52.86 & 0.092  & 0\\ \midrule 
Total & 575.80 & 1.0  & 1.0 \\ \bottomrule
\end{tabular}
\end{table}

\replaced[id=PSK]{Based on the state-level analysis, the analyst may consider that club stores are particularly relevant when assessing their role in the antitrust market. Investigative efforts could focus on exploring the extent to which consumers view the merging parties' stores and club stores as close substitutes.}{From the state-level analysis, the analyst can infer that only club stores merit attention regarding whether they are in the relevant antitrust market; investigative resources should be directed to examining whether consumers find merging parties’ stores and club stores as close substitutes.} In \ref{section:firm.level.analysis}, I apply the Shapley-Shubik power index to analyze the importance of each competitor firm in triggering the structural presumption. This exercise illustrates that the Shapley values and the Shapley-Shubik power index can be calculated with many players.

\subsection{Local Market Analysis}
Consumers mostly shop at supermarkets around residences or workplaces, as traveling is costly. As such, competition in the supermarket industry is local in nature. The FTC defined the relevant geographic markets as areas that range from a two- to ten-mile radius around each supermarket. As no further details are provided in the FTC's publicly released complaints, I simply define local ``circle'' markets as all grocery stores within a 5-mile radius around each of the defendant's supermarkets.\footnote{In other words, each Albertsons/Safeway store forms a 5-mile circle market that can be interpreted as a local geographic market. Defining local markets as a collection of stores around each ``center store'' is useful because the analyst does not need to define arbitrary geographic boundaries. If a local geographic market, however defined, does not have a high level of market concentration or significantly raise concentration, circle markets defined around center stores will likely not raise a presumption of illegality as well. In practice, antitrust authorities will consider specific geographic characteristics (e.g., rivers or highways) to determine the relevant geographic market and adjust the radius according to the urbanicity of local markets.} I calculate the post-merger HHI and $\Delta$HHI in each local market.

\begin{figure}[htbp!]
    \centering
    \includegraphics[scale=0.5]{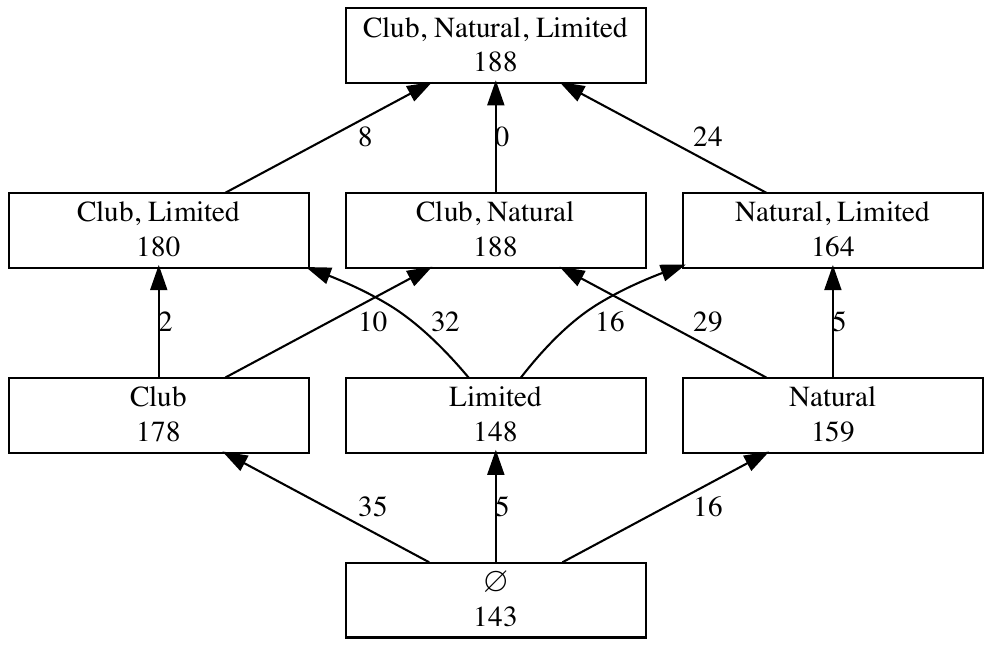}
    \caption{Number of Presumptive Local Markets}
    \label{figure:number.of.presumptive.local.markets}
\end{figure} 

Figure \ref{figure:number.of.presumptive.local.markets} shows the number of presumptive local markets. \replaced[id=PSK]{Out of 243 local markets (defined by 75 Albertsons and 168 Safeway stores in WA), t}{T}here are 143 presumptive markets \added[id=PSK]{(i.e., those that reach the merger guideline thresholds)} when all store formats are included in the market definition. However, the number increases to 188 when club, natural, and limited assortment stores are excluded from the market definition (only supermarkets and supercenters are considered), indicating 45 markets are sensitive to market definition. \replaced[id=PSK]{Notably, excluding club stores significantly raises the number of presumptive markets, highlighting the importance of carefully considering the role of club stores.}{Removing club stores from the market definition significantly raises the number of presumptive markets, indicating that club stores deserve major attention.}

\begin{figure}[htbp!]
    \centering
    \includegraphics[scale=0.45]{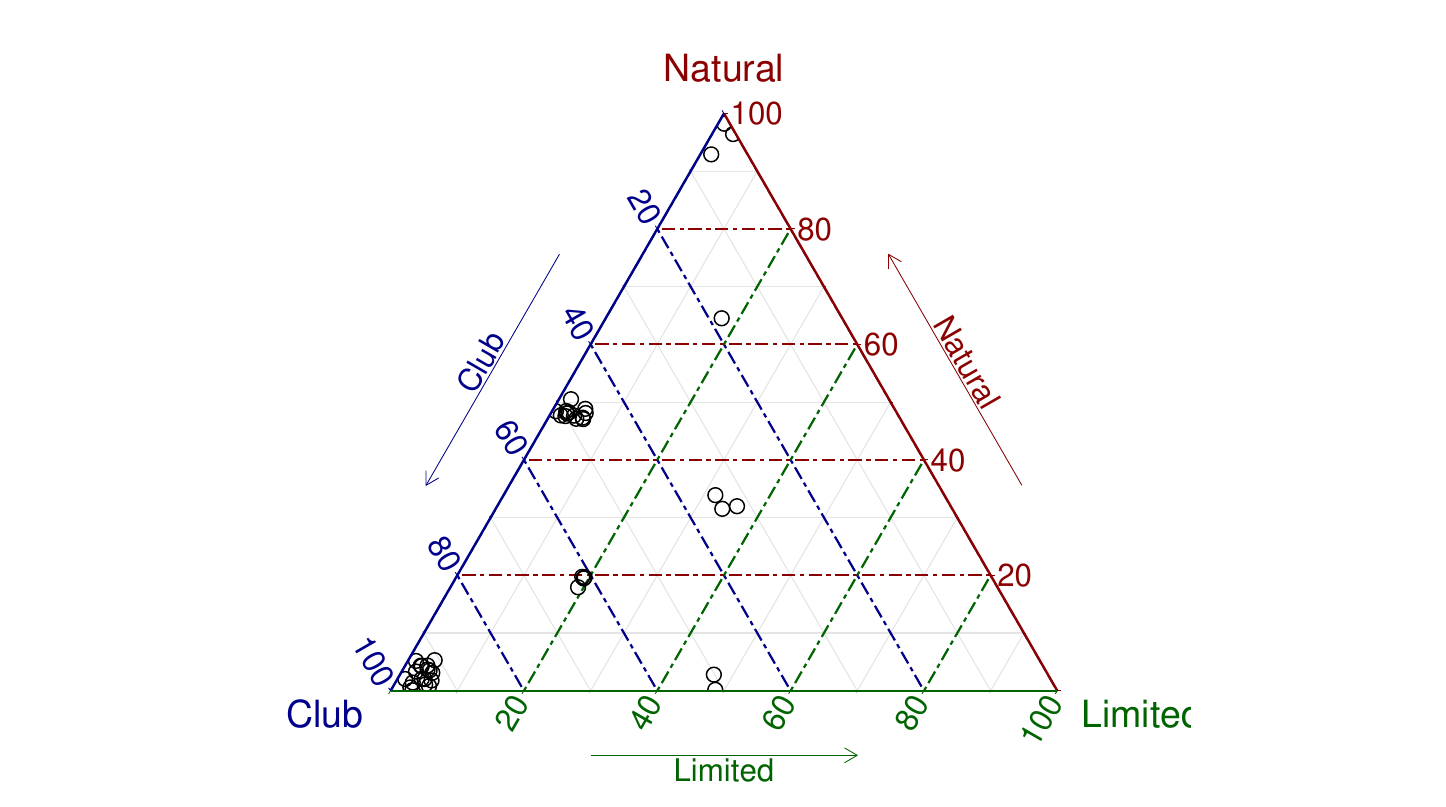}
    \caption{Shapley-Shubik Power Index of Local Markets}\label{figure:sspi.of.local.markets}
\end{figure}

Table \ref{table:local.market.sspi} reports the frequency of SSPI structure for the 45 sensitive markets. For example, out of the 45 sensitive markets, 18 markets have $(\text{SSPI}_{\text{Club}},\text{SSPI}_{\text{Natural}},\text{SSPI}_{\text{Limited}}) = (1,0,0)$, meaning whether the structural presumption is met only depends on the exclusion club stores (club format is the ``dictator'' in these local markets). The distribution of SSPIs across the local markets indicates that the club format tends to have a high power index. Figure \ref{figure:sspi.of.local.markets} visualizes Table \ref{table:local.market.sspi} (random noises were added to the points to enhance visibility). It also shows that clubs and natural stores have a high power index. In contrast, limited assortment stores are less important.

\begin{table}[htbp!]
\centering
\caption{Local Market SSPI \label{table:local.market.sspi}}
\begin{threeparttable}
\begin{tabular}{cccc} \toprule
\multicolumn{3}{c}{SSPI Structure} & \\ \cmidrule(lr){1-3}
Club & Natural & Limited & Frequency \\
\midrule
1.000 & 0.000 & 0.000 & 18\\
0.667 & 0.167 & 0.167 & 4\\
0.500 & 0.500 & 0.000 & 14\\
0.500 & 0.000 & 0.500 & 2\\
0.333 & 0.333 & 0.333 & 3\\
0.167 & 0.667 & 0.167 & 1\\
0.000 & 1.000 & 0.000 & 3\\ \midrule
0.663 & 0.274 & 0.063 & Mean \\ \bottomrule
\end{tabular}
\begin{tablenotes}
    \footnotesize
    \item \added[id=PSK]{\emph{Notes:} Table reports the frequency of SSPI distribution across 45 local markets (defined by Albertsons/Safeway stores) that are sensitive to market definition. The last row reports the overall mean of SSPI.}
\end{tablenotes}
\end{threeparttable}
\end{table}

The Shapley-Shubik power index shows that whether the structural presumption is met largely depends on whether club stores can be removed from the market definition. Exclusion of natural stores can also have a high impact, but limited assortment stores tend to matter little. In sum, the sensitivity analysis reveals which markets are sensitive to market definition, which formats are more important than others, and how investigative resources should be allocated to defend the conclusions on merger illegality.

\added[id=PSK]{In retail mergers, it is common to test the sensitivity of the analysis with respect to not only product market definitions but also geographic market definitions. In \ref{section:geographic.market}, I provide an example of how the analyst might use the proposed framework to incorporate geographic market considerations. Specifically, I assume the analyst also wants to test how the results would change as the radius of local markets is increased from 5 miles to 10 miles. Note, however, that market concentration statistics such as HHI need not be monotonic with respect to the local market radius. Whether expanding the market increases the HHI depends on whether there are relatively more stores owned by the merging parties than third-party competitors in the 5-10 miles band.}

\section{Conclusion \label{section:conclusion}}
Despite the debate on its economic significance, market definition continues to be a pivotal focus in antitrust cases.
Yet, attempts to pinpoint the proper market definition often prove fruitless. The FTC-DOJ Merger Guidelines acknowledge that a single case may be subject to multiple market definitions. Even in the most ideal case, the adversarial nature of litigation requires the plaintiffs, defendants, and court to scrutinize the sensitivity of the conclusion with respect to market definition.

Contrary to existing works that attempt to find the best approach for defining an antitrust market, I take the possibility of entertaining multiple market definitions as given and propose a simple framework for conducting an organized sensitivity analysis. I show that Hasse diagrams can help summarize the sensitivity of model outcomes with respect to market definition. Moreover, the Shapley value and the Shapley-Shubik power index are simple tools for quantifying the impact of including or excluding firms from the relevant market.

Although I have focused on a particular application, the proposed idea can be applied more broadly. For example, estimating economic margins is also an area of contention as margins are key inputs to competitive effects analysis; the plaintiffs and defendants may disagree on the cost categorization, which naturally leads to the need for sensitivity analysis. More generally, a researcher may use the proposed tools to evaluate the sensitivity of policy recommendations with respect to a set of discrete assumptions. \added[id=PSK]{A Hasse diagram may also be helpful in summarizing how the size of the partially identified sets with respect to alternative assumptions on players' information and solution concepts in discrete game models; see, e.g., \citet{koh2023stable}. In the field of empirical industrial organization, it is also standard to consider multiple market definitions or explore the implication of different nesting structures in nested logit models, which can be framed as a form of sensitivity analysis.}\footnote{\added[id=PSK]{I thank an anonymous referee for making this observation.}} Developing more economic tools that facilitate organized sensitivity analysis will be valuable. I leave further development and applications to future work.

%%% APPENDIX %%%
\appendix
% \part*{Appendix}
\renewcommand{\thesection}{Appendix \Alph{section}}

\section{Monotonicity of Popular Evaluation Metrics \label{section:monotonicity}}

Removing products or firms from the market narrows the market definition and increases the market shares of each remaining firm. I show that popular measures of market power and competitive effects that depend on market shares have the monotonicity property under the logit demand assumption. To set the stage, let $m$ denote a market that consists of a set of firms. If $m^1 \supseteq m^0$, then $m^1$ is a broader market. Let $x_j$ represent firm $j$'s sales. The share of firm $j$ is $s_{j,m} \equiv x_j / \sum_{l \in m} x_l$.

\subsection{Concentration Ratios}
The concentration ratio is calculated as the sum of the market share percentage held by the largest specified number of firms in an industry. Define the concentration ratio of a group $g \subseteq m$ as
\[
f(g;m) = \sum_{j \in g}s_{j,m}.
\]
If $m^1 \supseteq m^0$, then $s_{j,m^0} \geq s_{j,m^1}$ for all $j \in m^0$. It follows that $f(g;m^0) \geq f(g;m^1)$, i.e., defining a smaller market leads to a higher concentration ratio statistic.

\subsection{Upward Pricing Pressure}
In a single-product firm setting, \citet{farrell2010antitrust} defines the upward pricing pressure of firm $j$ merging with firm $k$ is defined as
\[
\mathit{UPP}_j = (p_k - c_k) \cdot D_{j \to k},
\]
where $(p_k - c_k)$ is the pre-merger price-cost margin of firm $k$ and $D_{j \to k}$ is the diversion ratio from $j$ to $k$. Under the logit demand assumption, diversion ratios are proportional to share:
\[
D_{j \to k} = \frac{s_k}{1 - s_j}.
\]
The numerator is increasing in $s_k$ and the denominator is decreasing in $s_j$. Thus, defining a smaller market increases the share of firms $j$ and $k$, leading to higher diversion ratio estimates. It follows that defining a smaller market leads to a higher estimate of upward pricing pressure.

\subsection{Compensating Marginal Cost Reductions}
\citet{werden1996robust} shows that the compensating marginal cost reduction for firm $j$ is
\[
\ddot{c}_j = \frac{m_j D_{j \to k} D_{j \to k} + m_k D_{k \to j} p_k / p_j}{(1 - m_j) (1 - D_{j \to k} D_{k \to j})},
\]
where $m_l = (p_l - c_l) / p_l$ is the pre-merger margin. Since the numerator is increasing in diversion ratios and the denominator is decreasing in diversion ratios, it follows that the compensating marginal cost reductions increases with smaller market definition under the logit assumption.

\subsection{Herfindahl-Hirschman Index}
The HHI of market $m$ is defined as
\begin{equation}\label{equation:HHI}
H_m = \sum_{j \in m} s_{j,m}^2.
\end{equation}
For a given market definition $m$ that includes the merging firms $a$ and $b$, the increase in HHI from the merger is
\begin{equation}\label{equation:Delta.HHI}
\Delta H_m \equiv H^\text{post}_m - H^\text{pre}_m = (s_{a,m} + s_{b,m})^2 - (s_{a,m}^2 + s_{b,m}^2) = 2s_{a,m} s_{b,m}.    
\end{equation}
From \eqref{equation:Delta.HHI}, it is apparent that defining a smaller market increases $\Delta H_m$.

However, establishing the monotonicity of HHI (in level) is more subtle because HHI depends on the share of all firms in the given market. It is well-known that HHI is decreasing in the number of firms when they are symmetric \citep{tirole1988theory}. However, HHI need not be decreasing in the number of firms when firms are asymmetric because including a large firm may result in a higher HHI.\footnote{To see this, consider the following example. Suppose that $m = \{1,2,3\}$, $x_a = 1$, $x_b = 1$, and $x_c = 98$. Consider $m' = \{1,2\}$ obtained by removing the largest firm $c$. Then $f(m) = 9,606$ but $f(m') = 5,000$, so a narrower market decreases HHI. More generally, driving the sale of the largest firm to infinity makes the HHI converge to the level of a monopoly market.} Nevertheless, defining a narrower market leads to a higher HHI insofar as the volume of the marginal firm is sufficiently small.

Consider 
\[
H_{m} = \frac{x_1^2 + ... + x_J^2}{(x_1 + ... + x_J)^2}.
\]
Suppose that $x_J$ is the sale of the marginal firm. I consider how $H_m$ behaves \added[id=PSK]{with respect to changes in $x_J$} when $x_J \approx 0$, which mimics \replaced[id=PSK]{an experiment of adding a small marginal firm $J$ in the market definition}{leaving $x_J$ out of the market but including it in the relevant market}. Taking partial derivative with respect to $x_{J}$ yields
\[
\frac{\partial H_m}{\partial x_J} = 2x_J(x_1 + ... x_J)^{-2} + (x_1^2 + ... + x_J^2)(-2)(x_1 + ... + x_J)^{-3}.
\]
The above is nonpositive if and only if
\[
x_J (x_1 + ... x_J) \leq (x_1^2 + ... + x_J^2).
\]
Thus, if $x_J \to 0$, the left-hand side of the inequality approaches zero, whereas the right-hand side remains strictly positive. This shows that \replaced[id=PSK]{$\partial H_m / \partial x_J \leq 0$}{$\partial H_m / \partial x_J$} when $x_j \approx 0$, establishing that HHI decreases as the relevant market includes smaller players at the margin.

\section{Firm-Level Analysis \label{section:firm.level.analysis}}
Again, using the \replaced[id=PSK]{AC Nielsen Trade Dimensions}{TDLinx} data from 2009, I also examine the impact of the top 9 chains in Washington State in triggering the structural presumption (holding Albertsons and Safeway's presence fixed). I assume that a market definition triggers a structural presumption if the post-merger HHI is above 1,800 and the change in HHI is above 100. Table \ref{table:empirical.application.shapley.shubik} reports the Shapley-Shubik power index.

\begin{table}[htbp!]
\centering
\caption{Firm-level Shapley-Shubik Power Index \label{table:empirical.application.shapley.shubik}}
\begin{tabular}{lcc} \toprule
Firm &  SSPI \\ \midrule
Kroger  &  0.313 \\
Costco  &  0.313 \\
Walmart   & 0.313 \\
Winco  &  0.018 \\
Trader Joe's  &  0.011 \\
The Markets  &  0.011 \\
Whole Foods  &  0.011 \\
Grocery Outlet  &  0.011 \\
Natural Grocers  &  0.000 \\ \midrule
Total  & 1.0 \\ \bottomrule
\end{tabular}
\end{table}

Walmart, Kroger, and Costco account for approximately 94\%. Surprisingly, Walmart, Kroger, and Costco have the same influence on triggering the structural presumption. Thus, the Shapley value reveals that a firm's importance in triggering the structural presumption is not strictly proportional to its total sales. Shapley values clarify how an outcome measure of interest depends on which firms/products are included/excluded in the market definition.

\section{Incorporating Geographic Market \label{section:geographic.market}}

\added[id=PSK]{In the main analysis, I focused my attention on sensitivity analysis with respect to product market definition while fixing the radius of geographic markets to be 5 miles around each center store. However, in retail merger cases, it is also common to test the sensitivity with respect to the definition of the geographic market.}\footnote{\added[id=PSK]{Note, however, that market concentration or unilateral effects statistics may not be monotonic with respect to geographic market radii. For instance, increasing the radius from 5 miles to 10 miles in my example may increase local market HHI if there are many Albertsons/Safeway stores in the 5-10 mile ring but decrease local market HHI if there are none.}} \added[id=PSK]{Here, I provide a simple example of incorporating geographic market considerations.}

\begin{figure}[htbp]
    \centering
    \includegraphics[width=0.5\linewidth]{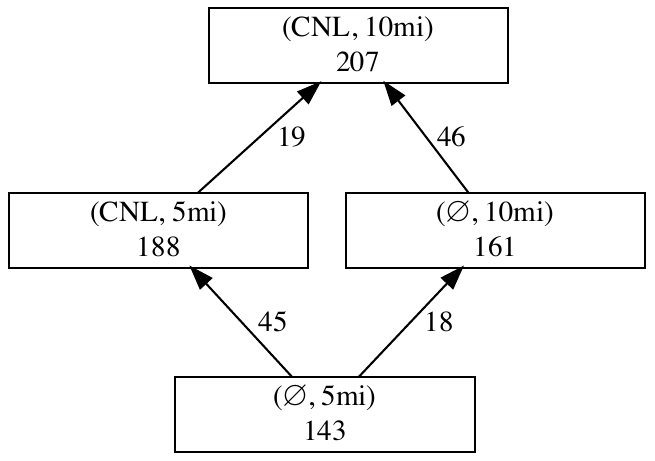}
    \caption{Sensitivity of Presumptive Market Counts to Product and Geographic Market Definitions}
    \label{figure:sensitivity.to.product.and.geographic.market.definitions}
\end{figure}

\added[id=PSK]{Suppose the analyst wants to test sensitivity in two dimensions. Let $x=(x_P,x_G)$ where $x_P = \emptyset$ if all formats are included in the product market definition and $x_P=\text{CNL}$ if clubs, natural, and limited assortment formats are excluded; $x_G = 5\text{mi}$ if the local market radius is set to 5 miles and $x_G = 10\text{mi}$ if the radius is set to 10 miles. Figure \ref{figure:sensitivity.to.product.and.geographic.market.definitions} plots the Hasse diagram that shows the number of presumptive local markets as a function of product and geographic markets. For example, setting the market definition as $(\emptyset,5\text{mi})$ (i.e., considering all grocery formats and defining local markets to be all stores within 5 miles of each center store, as done in my baseline analysis) leads to 143 presumptive markets. However, increasing the local market radius to 10 miles increases the count to 161. Excluding clubs, natural, and limited assortment stores from the product market definition increases the count by 46 to 207. }

\begin{table}[htbp]
    \centering
    \caption{Product and Geographic Market SSPI}
    \label{table:product.and.geographic.market.sspi}
    \begin{threeparttable}
    \begin{tabular}{lcc} \toprule
    Product & Geographic & Frequency \\ \midrule
        1.00 & 0.00 & 21  \\
        0.50 & 0.50 & 23 \\
        0.00 & 1.00 & 9 \\ \midrule
        0.61 &  0.39 & Mean \\ \bottomrule
    \end{tabular}    
    % \begin{tablenotes}
    %     \footnotesize
    %     \item 
    % \end{tablenotes}
    \end{threeparttable}
\end{table}

\added[id=PSK]{Table \ref{table:product.and.geographic.market.sspi} tabulates the distribution of SSPI for local markets do not hit the structural presumption threshold if the market definition $x=(\emptyset, 5\text{mi})$ but does so if $x=(\text{CNL},10\text{mi})$. The two-dimensional SSPI is calculated assuming ``participation'' means excluding clubs, natural, and limited assortment formats for the product market and increasing the local market radius from 5 miles to 10 miles for the geographic market, as implied by the Hasse diagram in Figure \ref{figure:sensitivity.to.product.and.geographic.market.definitions}. Overall, whether the local markets meet the structural presumption tends to be slightly more sensitive to product market definition than geographic market definition, as can be seen from the mean of SSPI in the bottom row of Table \ref{table:product.and.geographic.market.sspi}.}

\added[id=PSK]{Note that Table \ref{table:product.and.geographic.market.sspi} considers a total of 53 (=21+23+9) markets whereas Figure \ref{figure:sensitivity.to.product.and.geographic.market.definitions} shows a total increase of 64 (from 143 to 207). The counts can be different because HHI may be non-monotonic with respect to the local market radius. For example, let $y=(y_1,y_2,y_3)$ represent a vector of indicators for whether markets 1, 2, and 3 meet the structural presumption. Suppose $y=(1,0,0)$ when $x_G = 5\text{mi}$ (i.e., local markets 1 hits the threshold, but markets 2 and 3 do not). Since $y$ need not be monotonic with respect to the local market radius, increasing the radius to $x_G = 10\text{mi}$ may produce $y=(0,1,1)$, in which case the total count of presumptive markets has increased even though some markets that hit the structural presumption no long hit the threshold. }

%%% REFERENCE %%%
% \clearpage
\singlespacing
\bibliographystyle{apalike}
\bibliography{references}

\end{document}